\documentclass{amsart}
\usepackage[utf8]{inputenc}
\usepackage{amsmath}%
\usepackage{amsfonts}%
\usepackage{amssymb}%
\usepackage{graphicx}
\usepackage{xr}
\usepackage{multirow}
\usepackage{url}
\usepackage{caption, subcaption}
\usepackage{tabularx}
\usepackage{booktabs}
\usepackage{siunitx}

\numberwithin{equation}{section}
\newcolumntype{C}{>{\raggedleft\arraybackslash}X}

\begin{document}
\title[Comparison between continuous and discrete doses]{Comparison between continuous and discrete doses using Escalation With Overdose Control}
\author{Márcio Augusto Diniz}
\author{Mourad Tighiouart}
\author{André Rogatko}
\address[DMA, TM, RA]{Samuel Oschin Comprehensive Cancer Institute, \newline
\indent 8700 Beverly Boulevard, Los Angeles, CA 90048}
\email[DMA]{marcio.diniz@cshs.org}%
\urladdr{http://www.github.com/dnz.marcio}
\email[TM]{mourad.tighiouart@cshs.org}%
\email[RA]{andre.rogatko@cshs.org}%


\begin{abstract}
Although there is an extensive statistical literature showing the disadvantages of discretizing continuous variables, categorization is a common practice in clinical research which results in substantial loss of information.
A large collection of methods in cancer phase I clinical trial design establishes dose of a new agent as a discrete variable. A noteworthy exception is the Escalation With Overdose Control (EWOC) design, where doses can be defined either as continuous or as a grid of discrete doses.
A Monte Carlo simulation study was performed to compare the operating characteristics of continuous and discrete dose EWOC designs. Four equally spaced grids with different interval lengths were considered. The loss of information was measured by several operating characteristics easier for clinicians to interpret, in addition to the usual statistical measures of bias and mean squared error.
Based on the simulations, if there is not enough knowledge about the true MTD value as commonly happens in phase I clinical trials, continuous dose scheme arises as an attractive option. 
\end{abstract}
\maketitle

\section{Introduction}

Measurements of continuous variables are made in all fields of medicine. In medical research such continuous variables are often converted into categorical variables by grouping values into two or more categories in order to have easier interpretations.

Cox \cite{c1957} derived an optimization criteria for discretizing a continuous covariable and showed that if a variable is normally distributed then categorizing it into six groups implies a minimum loss of 5.8\% of information, when using a quadratic loss function. The minimum loss is 7.99\% for five groups, 11.75\% for four groups and 19.02\% for three groups. Following the rationale of Cox, Connor \cite{c1972} found other criteria and reached similar conclusions based on the asymptotic relative efficiency of the Cochran-Armitage trend test \cite{a1955} when a continuous covariable, that is linearly related to a binary response variable, is discretized. In this way, several authors have pursued methodologies to provide optimal criteria of discretization for continuous covariables based on other test statistics. See \cite{ms1982}, \cite{wn1991}, \cite{ls1992}, \cite{shs1997}, \cite{msb2003}. 

On the other hand, extensive statistical literature (see \cite{c1983}, \cite{l1988}, \cite{a1994}, \cite{w1995}, \cite{mzpr2002}, \cite{im2003} and \cite{ras2006}) has advised against categorization due the loss of power and precision of the estimated quantities. In particular, Lagakos \cite{l1988} studied the effects of mis-measuring covariables, considering a likelihood test for a logistic model when a continuous covariable following Normal, Exponential or Uniform distribution, is categorized. He also evaluated the loss of information when the levels of an ordered categorical covariable is selected incorrectly considering the Cochran-Armitage trend test. 

This debate is not different for cancer phase I clinical trials, although with its peculiarities. Phase I trials represent the first testing of an investigational agent in humans and act as a point of translation of years of laboratory research into the clinic. The major objective in phase I trials is to identify a maximum tolerable dose (MTD) for subsequent studies, whereas the primary goal in phase II and III trials is treatment efficacy. While phase I trials in other areas of medicine enroll healthy participants, phase I oncology trials typically enroll patients who have advanced form of cancer and who have exhausted standard treatment options. 

Ideally, from a therapeutic perspective, clinical trials should be designed to maximize the number of patients receiving an optimal dose. The fundamental conflict underlying the design of cancer phase I clinical trials is that increasing the dose slowly to avoid unacceptable toxic events must be balanced against treating many patients at suboptimal or nontherapeutic doses. \cite{rsjt2007}

Traditionally, dose finding has been conducted according to the 3 + 3 principle and its variants, which was first described by Dixon and Mood \cite{dm1948} and require a pre-specified set of discrete doses. Although the use of rule-based designs is still prevailing, model-based designs such as the continual reassessment method (CRM) introduced by O`Quigley \cite{qpf1990} and Escalation With Overdose Control (EWOC) by Babb et al. \cite{brs1998} have been gaining popularity in clinical practice. \cite{vhvbs2016}

In these model-based designs, a parametric model is considered to describe the relationship between the probability of a dose limiting-toxicity (DLT) and the dose level of the new agent, which can be considered as a continuous or a discrete covariable.  

Clinical trials using continuous dose \cite{cblarefsfa2004, bahflrhdlc2009} are less often performed since clinicians are used to the up-and-down approach, although intravenous drugs are still more prevalent than oral drugs \cite{hj2010, wbbejkllmn2008} and continuous dose does not require to pre-specified a set of doses which is usually an arbitrary decision. 

Following the up-and-down approach, a large collection of methods establishes the dose of a new agent only as a discrete variable based on a pre-specified set of doses. However, Hu et al. \cite{hbj2013} and Chu et al. \cite{cpy2016} pointed out the common assumption by almost all existing methods that one of the pre-specified doses is the MTD does not hold in practice since often limited prior information regarding the dose-toxicity relationship of the experimental drug is available. 

A noteworthy exception is the EWOC design where doses can be defined either as continuous or as a grid of discrete doses. For discrete doses, EWOC still will consider the same continuous dose algorithm, but a final rounding step is added to the algorithm.

EWOC was the first dose-finding procedure to directly incorporate the ethical constraint of minimizing the chance of treating patients at unacceptably high doses. Its defining property is that the expected proportion of patients treated at doses above the MTD is equal to a specified value $\alpha$, the feasibility bound. \cite{tcr2012} 

The feasibility bound could varying along the trial as discussed by Tighiouart and Rogatko \cite{tr2010}. The rationale behind this approach is that uncertainty about the MTD is high at the onset of the trial and a small value of $\alpha$ offers protection against the possibility of administering dose levels much greater than the MTD. As the trial progresses, uncertainty about the MTD declines and the likelihood of selecting a dose level significantly above the MTD become significantly smaller. Chu et al. \cite{cpl2009} compared the performance of different versions of CRM with EWOC both constant and varying $\alpha$.

In this work, a Monte Carlo simulation study to compare the operating characteristics of continuous and discrete dose EWOC designs is presented. The loss of information will be evaluated using the statistical measures bias and mean squared error as well as specific measures for phase I clinical trials to quantify safety and efficacy of the trial. 

This article is organized as follows. In Section \ref{sec:ewoc}, the EWOC design is briefly introduced. The simulation study is described in Section \ref{sec:simulation} and its results are presented in Section \ref{sec:results} with discussion in Section \ref{sec:discussion}.

\section{Escalation with Overdose Control} \label{sec:ewoc}

In this section, the EWOC design is briefly reviewed as Babb et al. \cite{brs1998}. Let $X_{min}$ and $X_{max}$ denote the minimum and maximum dose levels available for use in the trial. Note that the dose given to the first cohort of patients is not necessarily equal to $X_{min}$ but there must be strong evidence that it is a safe dose. While the maximum dose is not a dose that one would ever use to treat a patient, it is a boundary that would never be exceeded. 

In this way, the minimum and maximum doses are respectively the lower and upper bound of the support of the MTD $\gamma$ which is defined by
\begin{eqnarray}
P(DLT| dose = \gamma) = \theta,
\label{eq:mtd}
\end{eqnarray}
such that $\theta$ is defined as the proportion of expected patients  to experience a medically unacceptable, dose-limiting toxicity if the MTD $\gamma$ is administered. 
The relationship between toxicity and dose level could be defined as
\begin{eqnarray}
P(DLT| dose = x) = F(\beta_0 + \beta_1x),
\label{eq:dlt}
\end{eqnarray}
where $F$ is a specified distribution function and $\beta_0$, $\beta_1$ are unknown parameters such that $\beta_1 > 0$. Following (\ref{eq:mtd}) and (\ref{eq:dlt}), the MTD will be given by
\begin{eqnarray}
\gamma 
&=& \frac{F^{-1}(\theta) - \beta_0}{\beta_1} \nonumber\\
&=& X_{min} + \frac{F^{-1}(\theta) - F^{-1}(\rho_0)}{\beta_1}, \nonumber
\end{eqnarray}
where $\rho_0$ denotes the probability of a DLT at the initial dose often established as $x_1 = X_{min}$. Using the definition of the MTD and probability of toxicity at initial dose, one can show that
\begin{eqnarray}
\beta_0 &=& \frac{\gamma F^{-1}(\rho_0) - X_{min}F^{-1}(\theta)}{\gamma - X_{min}}, \nonumber\\
\beta_1 &=& \frac{F^{-1}(\theta) - F^{-1}(\rho_0)}{\gamma - X_{min}}.
\label{eq:beta}
\end{eqnarray}
Denoting by $y_i$ the toxicity response (1 for DLT and 0 for no DLT) of the $ith$ patient. The likelihood of the data $D_k = \{(x_i, y_i), u = 1, \ldots, k\}$ after observation of $k$ patients is
\begin{eqnarray} 
L(\rho_0, \gamma|D_k) = \prod_{i = 1}^k F(\beta_0 + \beta_1x_i)^{y_i}[1 - F(\beta_0 + \beta_1x_i)]^{1 - y_i}.
\end{eqnarray}
for $(\beta_0, \beta_1)$ defined as funtions of $(\rho_0, \gamma)$ given in (\ref{eq:beta}). 

Prior information is incorporated for $(\rho_0, \gamma)$ such that prior distributions could be chosen under the restrictions of  $\gamma \in [X_{min}, X_{max}]$ and $\rho_0 \in (0, 1)$. The natural choice is a $Beta(a_\rho, b_\rho)$ distribution for $\rho_0$ and a re-escaled $Beta(a_\gamma, b_\gamma)$ distribution for $\gamma$, but Tighouart et al. \cite{trb2005} examine a large class of prior distributions which could be considered. 

Finally, the calculation of the posterior distribution for $(\rho_0, \gamma)$ can be evaluated \cite{rtcl2016} and implemented  using numerical integration and Markov chain Monte Carlo sampler
\begin{eqnarray}
\pi(\rho_0, \gamma|D_k) = c(D_k)L(\rho_0, \gamma|D_k)\pi(\rho_0, \gamma),
\end{eqnarray}
where $c(D_k)$ is a normalizing constant. 

Hence, the $k + 1$ patient receives the dose given by the $\alpha$-quantile of the $\gamma$ posterior distribution
\begin{eqnarray}
x_k = \Pi^{-1}(\alpha | D_k),
\end{eqnarray}  
for $\alpha$ being the probability that the dose selected by EWOC is higher than the MTD. 

There are several suggestions for the choice of the feasibility boundary. Originally Babb et al. \cite{brs1998} suggested a fixed feasibility boundary $\alpha$ equal to 0.25, denoted by F(0.25). Babb and Rogatko \cite{br2001} suggested an increasing feasibility boundary until to 0.5 with initial $\alpha$ equal 0.25, denoted by I(0.25, 0.05). Rogatko and Tighiouart \cite{rt2013} suggested a similar strategy, but conditionally to the previous patient has no DLT, denoted by C(0.25, 0.05). 

For a discrete set of doses, EWOC is performed using dose as continuous with an additional step that could either round down $x_k$ to the closest dose  prioritizing safety or round to the nearest dose preferring ability to explore the available grid of doses. 

\section{Simulation study} \label{sec:simulation}

Escalation with Overdose Control was applied for continuous dose and discrete dose which were denoted by dose schemes. The minimum and maximum doses were standardized as $X_{min} = 0$ and $X_{max} = 1$. Considering the discrete dose scheme, four equally spaced grids with different interval lengths between two doses given by 0.05, 0.10, 0.2 and 0.25 such that each grid will be indicated by its interval length. The grid 0.05 has 21 doses, the grid 0.10 has 11 doses, the grid 0.20 has 6 doses and the grid 0.25 has 5 doses. 

The DLT proportion threshold was established equal to 0.33 such that four true values of MTD $= \{0.2, 0.4, 0.6, 0.8\}$ and four true distributions Logistic(0, 1), Normal(0, 2), Skew-Normal(0, 2, 3) and Skew-Normal(0, 2, -3) were considered. Three different sample sizes $n = 20, 40, 60$ with cohorts of one patient were treated. In the discrete dose scheme, rounding nearest approach was applied and skipping doses was not allowed.

A Monte Carlo study was performed considering 1000 replicates for each study design. The classical statistical measures of bias and mean square error (MSE) were evaluated as well as operating characteristics more interpretable for clinicians: average DLT rate, percentage of trials which DLT proportion is outside the target rate interval $\theta \pm 0.10$, the percentage of trials with the estimated MTD within the optimal MTD interval defined as $\mbox{True MTD} \pm 0.15 \times \mbox{True MTD}$ and the optimal toxicity interval defined as $\theta \pm 0.10$, the percentage of patients receiving optimal doses defined by the optimal MTD and toxicity intervals. 

From the perspective of a patient participating in a dose finding trial, the best design is the one with the highest proportion of patients receiving optimal doses such that optimal doses could be defined in different perspectives. In this way, it is worth to characterize the discrete dose schemes based on the number of possibles doses could be considered as optimal doses based on the optimal MTD interval which can be seen in Table \ref{tab:n_optimal_doses01}. 

\begin{table}[h]
	\centering
		\begin{tabular}{cccccc}
		\hline
		\multirow{2}{*}{True MTD} & \multirow{2}{*}{Optimal Interval} & \multicolumn{4}{c}{dose scheme}\\
		\cline{3-6}
			 &  & D0.05 & D0.10 & D0.20 & D0.25 \\	
		\hline
		0.2 & (0.17 ; 0.23) & 4.8  (1) & 9.1  (1)  & 16.7 (1) & 0.0  (0)\\
		0.4 & (0.34 ; 0.46) & 14.3 (3) & 9.1  (1)  & 16.7 (1) & 20.0 (1)\\
		0.6 & (0.51 ; 0.69) & 14.3 (3) & 9.1  (1)  & 16.7 (1) & 0.0  (0)\\
		0.8 & (0.68 ; 0.92) & 23.8 (5) & 27.3 (3)  & 16.7 (1) & 20.0 (1)\\
		\hline
		\end{tabular}
		\caption{Percentage (Number) of doses relative to each discrete dose scheme which can be considered as an optimal dose based on the optimal MTD interval ($\mbox{True MTD} \pm 0.15 \times \mbox{True MTD}$)}
		\label{tab:n_optimal_doses01}
\end{table}

Dose schemes D0.05 and D0.10 for some true MTD values present more than one dose which can be considered as an optimal dose. Furthermore, the dose scheme D0.25 does not contain any possible dose which could be considered as an optimal dose for the true MTD values 0.2 and 0.6. 

On the other hand, slight differences from the true MTD could imply notable differences on toxicity, therefore optimal doses can also be defined based on the optimal toxicity interval. The dose schemes are evaluated based on the optimal toxicity interval in Table \ref{tab:n_optimal_doses02}. All discrete dose schemes contains at least one optimal dose on the perspective of the optimal toxicity interval such that all discrete dose schemes contain at least 30\% of the doses which can be considered as optimal doses when the true MTD is 0.6 and 0.8. If the toxicity optimal interval is defined as $\theta \pm 0.05$ instead of $\theta \pm 0.10$, then the results for optimal toxicity interval are similar to optimal MTD interval. 

\begin{table}[h]
	\centering
		\begin{tabular}{ccccc}
		\hline
		\multirow{2}{*}{True MTD} & \multicolumn{4}{c}{dose scheme}\\
		\cline{2-5}
			 &  D0.05 & D0.10 & D0.20 & D0.25 \\	
		\hline
		0.2 & 14.3 (3) & 9.1 (1) & 16.7 (1) & 20.0 (1) \\
		0.4 & 23.8 (5) & 27.3 (3) & 16.7 (1) & 20.0 (1) \\
		0.6 & 38.1 (8) & 36.4 (4) & 33.3 (2) & 40.0 (2) \\
		0.8 & 47.6 (10) & 45.5 (5) & 50.0 (3) & 40.0 (2) \\
		\hline
		\end{tabular}
		\caption{Percentage (Number) of doses relative to each discrete dose scheme which can be considered as an optimal dose based on the optimal toxicity interval ($\theta \pm 0.10$) }
		\label{tab:n_optimal_doses02}
\end{table}

All the simulations were performed using the R-package EWOC available at GitHub \cite{d2017}. The following section will address several aspects about continuous and discrete dose schemes under different aspects of EWOC. 

\section{Results} \label{sec:results}

The working model is the Logistic model. The feasibility strategy C(0.05, 0.05) was fixed. There are 12 (4 true MTD $\times$ 3 sample sizes) simulations for each dose scheme and true model. The operating characteristics are presented in Figures 1 - 4 using median, 25\% and 75\% quantiles. 

\begin{figure}[h]
\caption{Bias and Square root of MSE as a function of true distribution and dose scheme}
		\includegraphics[scale = 0.7]{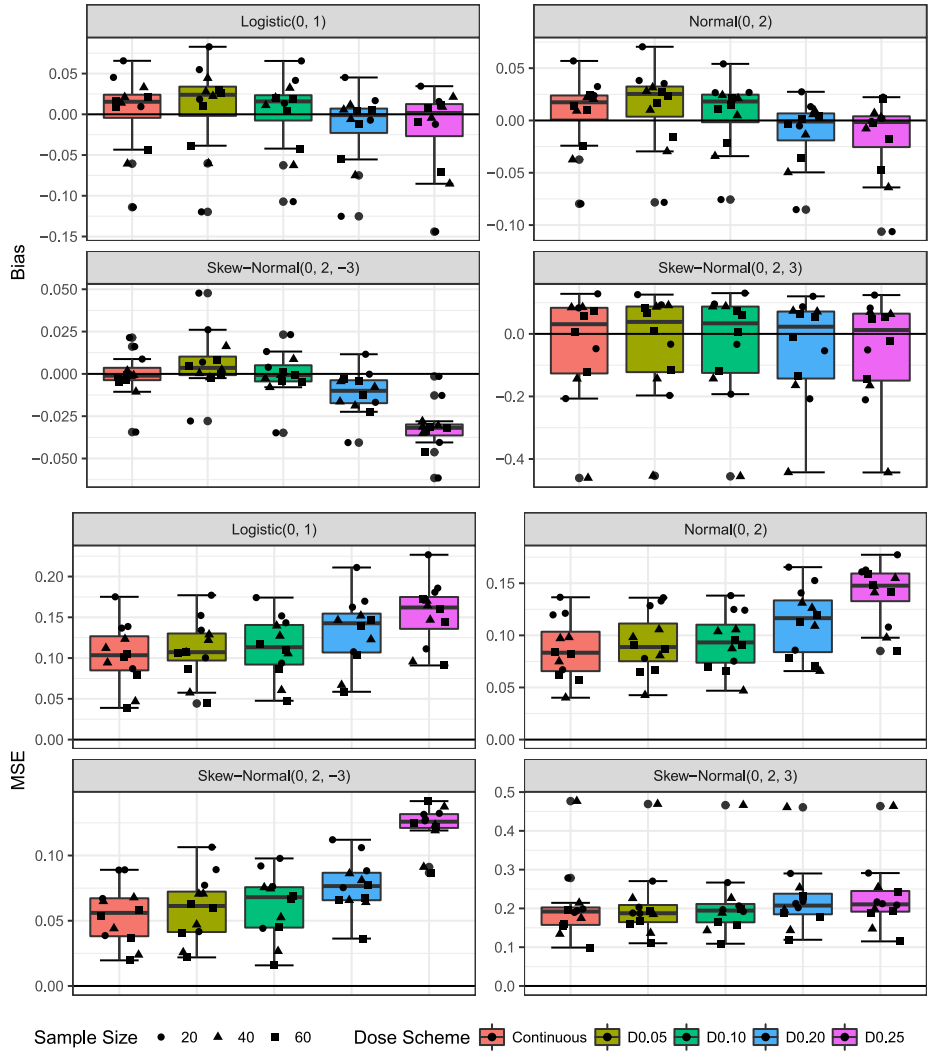}
\label{fig:fig1}
\end{figure}

Figure \ref{fig:fig1} shows results of the relative loss for the bias and the mean square error of the four discrete dose schemes. There is no significant differences among the dose schemes for bias, but there is a increasing pattern for the mean square error as the number of doses in the dose scheme decreases. 

\begin{figure}[h]
\caption{DLT Average and Percentage of trials such that the observed DLT probability is outside the interval $[\theta - 0.1;  \theta + 0.1]$ as a function of true distribution and dose scheme}
		\includegraphics[scale = 0.7]{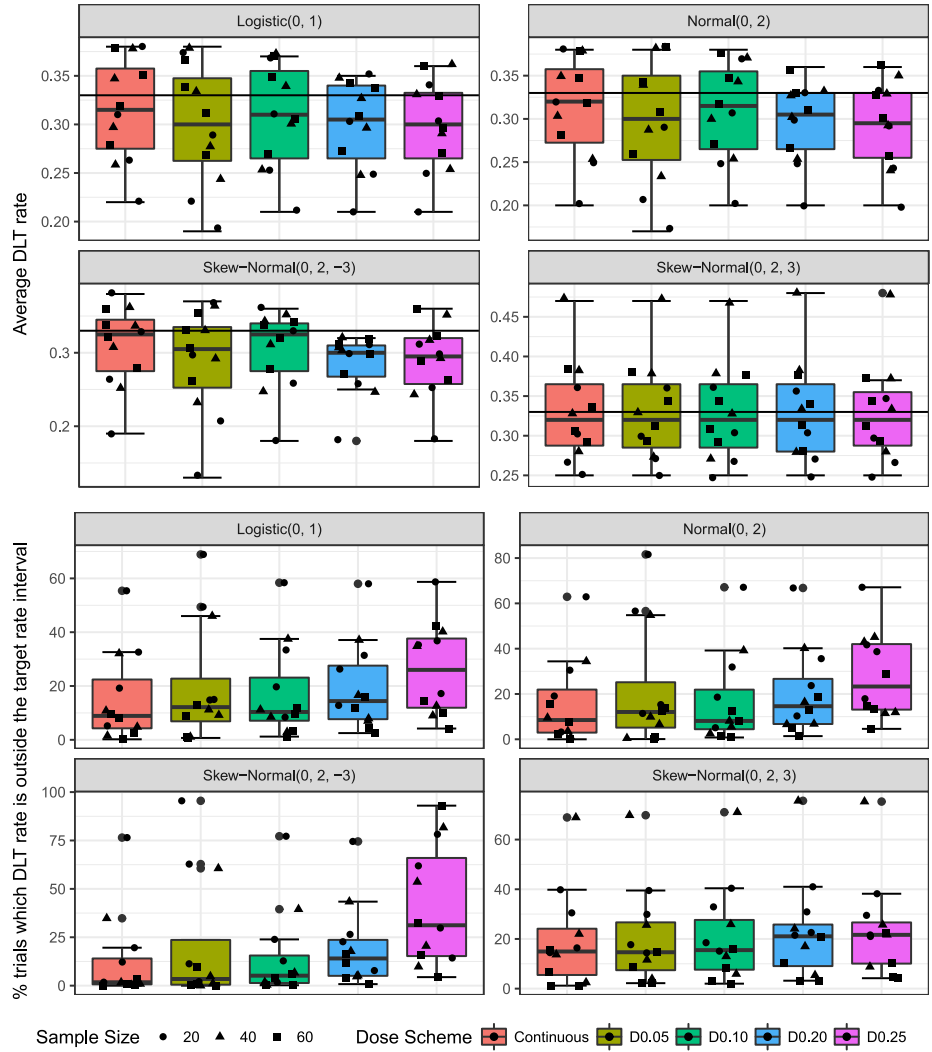}
\label{fig:fig2}
\end{figure}

The relative loss of information for the average DLT proportion and the percentage of trials which DLT proportion is outside the target rate interval are presented in Figure \ref{fig:fig2}. All discrete dose schemes are safer or equivalent to the continuous dose. However, the percentage of trials which DLT proportion is outside the target rate increases as the number of doses decreases indicating the lack of ability to explore the possible doses.

\begin{figure}[h]
\caption{Percentage of trials with the estimated MTD inside the MTD optimal interval and average percentage of patients receiving doses inside the optimal MTD as a function of true distribution and dose scheme}
	\centering
		\includegraphics[scale = 0.7]{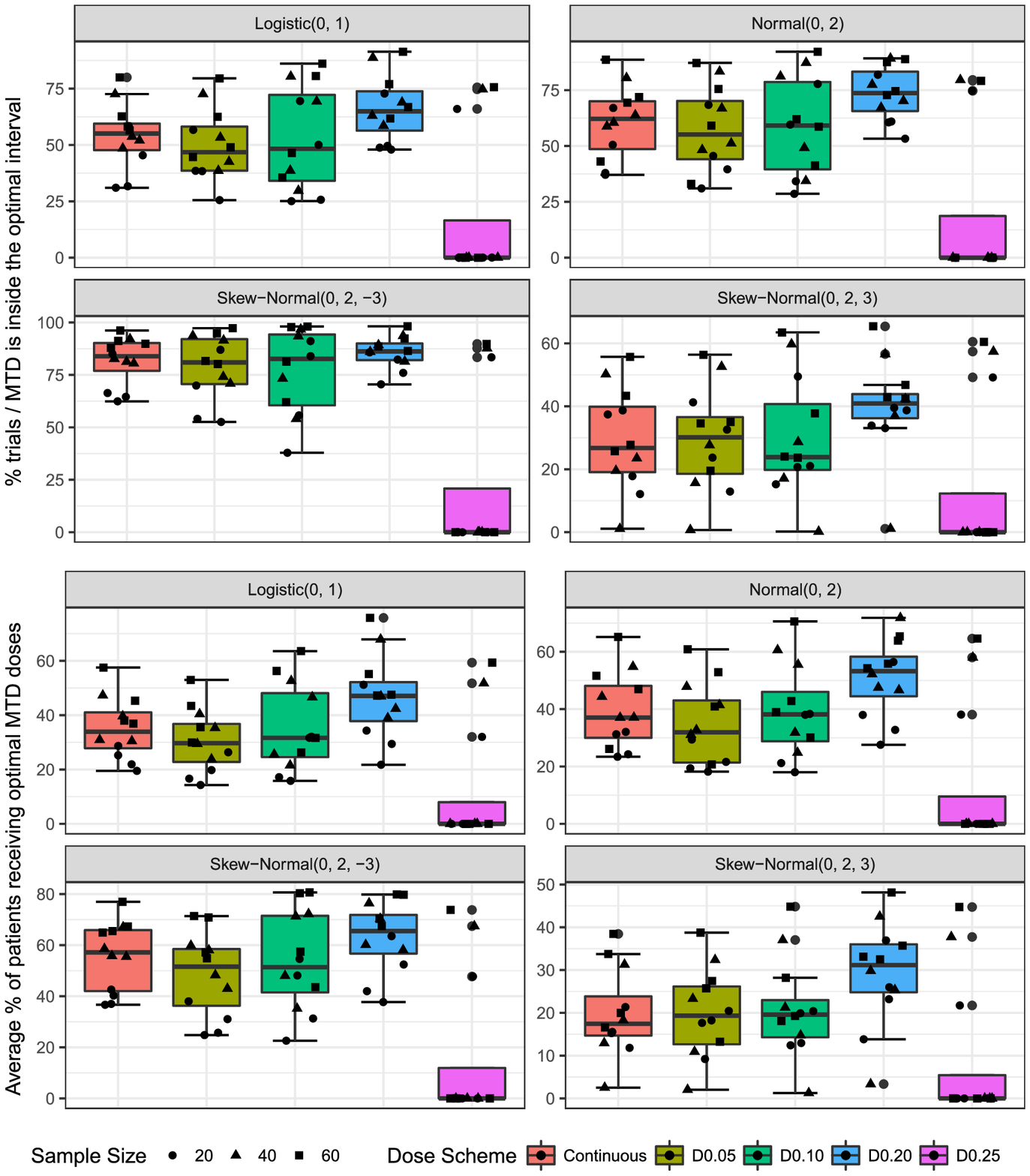}
\label{fig:fig3}
\end{figure}

Figure \ref{fig:fig3} presents the relative loss for the percentage of trials which the estimated MTD is inside the optimal MTD interval and the average percentage of patients receiving optimal MTD doses. The dose scheme D0.25 presents remarkably worse performance, as expected based on Table \ref{tab:n_optimal_doses01}, than the continuous dose and the other discrete dose schemes which contain the true MTD value for both operating characteristics. On the contrary, the dose scheme D0.2 surpasses the continuous dose since this dose scheme is the smallest set which always contains the true MTD among the discrete dose schemes considered.

\begin{figure}[h]
	\centering
	\caption{Percentage of trials with the estimated MTD  inside the toxicity optimal interval and average percentage of patients receiving doses inside the optimal MTD as a function of true distribution and dose scheme}
		\includegraphics[scale = 0.7]{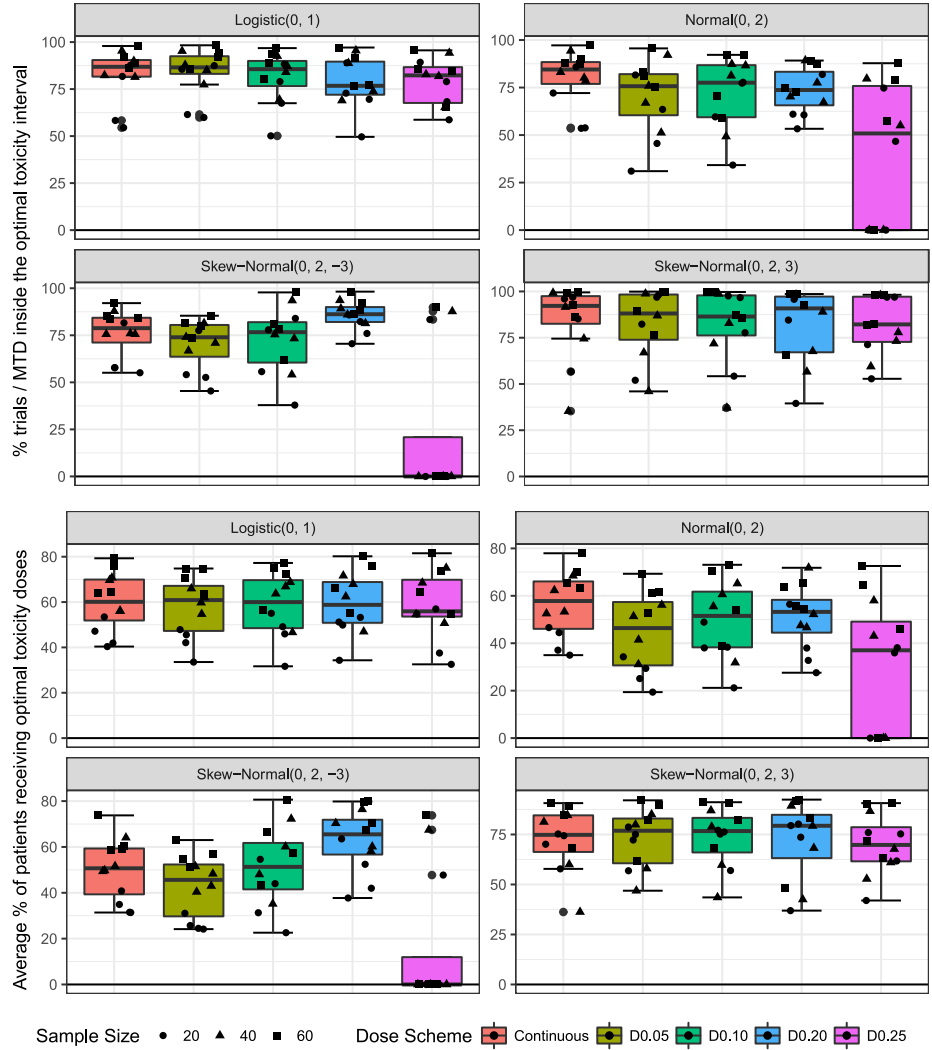}
\label{fig:fig4}
\end{figure}

The estimated MTD is inside the optimal toxicity interval and the average percentage of patients receiving optimal toxicity doses are presented in Figure \ref{fig:fig4}. Results are similar to Figure \ref{fig:fig3}, but less extent since former operating characteristics are less stringent as showed in Table \ref{tab:n_optimal_doses02}.

\section{Discussion} \label{sec:discussion}

Escalation with Overdose Control can be implemented based on either a continuous or a discrete dose scheme allowing the comparison between them. This work compared such dose schemes considering several misspecified scenarios and samples size with a conditional feasibility strategy. The dose schemes were evaluated based on the most common statistical, safety and efficiency measures of five doses schemes corresponding to a continuous dose and four different pre-specified set of doses with different increment sizes. 

The discretization of the continuous doses increases the variance of the MTD estimate and decreases the ability to explore the dose range. Moreover, the continuous dose scheme is favored based on the percentage of trials with the estimated MTD inside the optimal MTD and toxicity (defined as $\theta \pm 0.05$) intervals as well as the average percentage of patients receiving doses inside the optimal MTD and toxicity (defined as $\theta \pm 0.05$) intervals when the discrete dose scheme does not contain the true MTD or the pre-specified set of doses contains a large number of doses. 

Phase I clinical trials usually are performed with 5 or 6 doses chosen based on arbitrary criteria. Recent works \cite{hbj2013}, \cite{cpy2016} presented ideas to add new doses along the trial into the pre-specified discrete dose scheme. If there is not enough knowledge about the true MTD value as commonly happens in phase I clinical trials, continuous dose scheme arises as an attractive and more genuine option to avoid protocol amendments.

\newpage

\newpage

\newpage

\newpage

\section{Acknowledgement}
Supported in part by the NIH National Center for Advancing Translational Science (NCATS) UCLA CTSI Grant Number, Grant UL1 TR001881-01 (A.R. and M.T.), Grant R01 CA188480-01A1 (M.T. and A.R.). 

\bibliographystyle{plain}
\bibliography{paper} 

\end{document}